\documentclass[12pt]{article}

\usepackage{graphicx}

\newcommand{\be}{\begin{enumerate}}
\newcommand{\ee}{\end{enumerate}}
\newcommand{\beq}{\begin{equation}}
\newcommand{\eeq}{\end{equation}}
\newcommand{\beqa}{\begin{eqnarray}}
\newcommand{\eeqa}{\end{eqnarray}}

\newtheorem{def.}{Def.}

\begin{document}

\title{Generalized observers and velocity  \\
measurements in General Relativity}

\author{{\bf Paulo Crawford} $^{\rm 1,2 \footnote{Email address:crawford@cosmo.fis.fc.ul.pt}}$
 and {\bf Ismael Tereno} $^{\rm 1,3\footnote{Email address:ismael@cosmo.fis.fc.ul.pt}}$ \\
\small{1. Departamento de Fisica da Universidade de Lisboa} \\
\small{2. Centro de Astronomia e Astrofisica da Universidade de Lisboa}\\
 \small{3. Institut d'Astrophysique de Paris }
\date{}}
\maketitle

\baselineskip 24pt

\newpage

\begin{abstract}

\baselineskip 24pt

To resolve some unphysical interpretations related to velocity
measurements by static observers, we discuss the use of
generalized observer sets, give a prescription for defining the
speed of test particles relative to these observers, and show
that, for any locally inertial frame, the speed of a freely
falling material particle is always less than the speed of light
at the Schwarzschild black hole surface.

\end{abstract}

\vspace{1.2cm}

\newpage

\section{Introduction}

\baselineskip 20pt

The radial motion of a test particle falling in a Schwarzschild
black hole was treated by several authors \cite[p.298]{landau},
\cite[p.93]{Z&N},\cite[pp.19,20]{fronov},\cite[p.342]{shapiro},
\cite{cavspin1,cavspin2} who reached the same conclusion that the
particle velocity $v$ approaches the light velocity as the test
particle approaches the surface of the black hole, namely the
locus $r=2m$ (with a suitable choice of units), also known as the
event horizon or Schwarzschild radius. All these authors have in
common the use of observers whose worldlines are the integral
curves of a hypersurface orthogonal Killing vector field, that is,
static observers (called shell observers in \cite[p.2-33]{T&W})
and, as such, at rest with respect to the mass creating the
gravitational field. For example, Zel'dovich and Novikov say that
the velocity they use ``has direct physical significance. It is
the velocity measured by an observer who is at rest ($r, \theta,
\phi,$ constant) at the point the particle is passing."
\cite[p.93]{Z&N}. The particle's motion here is referred to the
Schwarzschild coordinate system in which the line element takes
the form
\begin{equation}
d\tau^2=\left( {1-{{2m} \over r}} \right)dt^2-\left({1-{{2m}\over r}}
\right)^{-1}dr^2-r^2\left( {d\theta^2 +\sin^2 \theta d\varphi ^2} \right),
\label{metschw}
\end{equation}
in geometric units ($c=G=1$).

Following along the same lines, Frolov and Novikov recently
\cite[pp.19,20]{fronov} add that ``The physical velocity $v$
measured by an observer who is at rest in the Schwarzschild
reference frame situated in the neighborhood of the freely moving
body is
\begin{equation}
v=\frac{dx}{d\tau}=\left[{\frac{g_{11}}{|g_{00}|}}\right]^{1/2}
\frac{dr}{dt} =\pm \frac{[E^2-1+r_g/r]^{1/2}}{E}. \label{velfn}
\end{equation}
If the falling body approaches $r_g(=2m)$, the physical velocity
$v=dx/d\tau$ constantly increases: $v\rightarrow 1$ as
$r\rightarrow r_g$." In Eq.(\ref{velfn}), $E=(1-2m/r)dt/d\tau$ is
a constant of motion which we may interpret for timelike geodesics
as representing the total energy per unit rest mass of a particle
following the geodesic in question, relative to a static observer
at infinity \cite[p.139]{wald}.

In their very well known textbook \cite[p.342]{shapiro}, Shapiro
and Teukolsky also produce a similar statement: ``\ldots the
particle is observed by a local static observer at $r$ to approach
the event horizon along a {\em radial} geodesic at the {\em speed
of light} \ldots"

All these statements have contributed to the wrong and widespread
view \cite{mitra1} that makes its way into the literature
\cite{mitra2}, of a test particle approaching the event horizon at
the speed of light {\em for all observers}, and not as a limiting
process for a static observer sitting at $r$, as $r\rightarrow
2m$. At the first sight, this view seems quite logical since we
expect the particle to cross the black hole surface in a finite
proper time. And if one accepts that a particle has the speed of
light with respect to a static observer (at $r=2m$), using locally
the velocity composition law from special relativity, he (or she)
concludes that the particle has the same speed of light with
respect to all observers. This is certainly something that
conflicts with the physical observation that, in a vacuum, no
material particle travels as fast as light. This has been very
nicely explained by Janis who established that the test particle
does indeed cross $r=2m$ with a speed less than the speed of light
\cite{janis2}. Here we take a similar view, and go one step
further in obtaining a general expression for geodesic radial
observers in terms of the constants of motion of both observer and
test particle.

In Sec. 2 we discuss the different mathematical status of
coordinate charts and reference frames, and compare this present
attitude with the early days when it was quite common the use of
`curvilinear four-dimensional coordinate system' and `frame of
reference' interchangeably. In Sec. 3, we review some standard
results and definitions of reference frames and observer sets, and
give a prescription for the speed at some space-time point
relative to generalized observers at that point. We find the speed
of any material particle to be strictly less than the speed of
light. In Sec. 4 we apply this general prescription to the
Schwarzschild field and reproduce Eq.(\ref{velfn}) for shell
observers, then we recall that there is a static limit, and we
obtain an expression, valid at $r=2m$, for the test particle's
square speed as a function of the constants of motion of the
observer and the particle, when both follow radially inward
geodesics. Finally, we give a brief discussion of the results.

\section{Coordinate Systems and Reference Frames}

One of the underlying principles of general relativity is the
freedom of choice of coordinates in the mathematical description
of laws and physical quantities. Indeed, the outcome of physical
measurements depends, in general, on the reference frame, that is,
on the `state of motion' of the observer, but cannot depend on
the coordinate system chosen, which may be completely arbitrary
and should be selected for convenience in the intermediate
calculation. Of course, certain coordinates may be preferred over
other coordinates in the sense that they are simpler or better
adapted to the symmetries of the gravitational field under
consideration.

The association of an arbitrary coordinate system with an
arbitrary frame of reference became standard in the literature for
many decades after the advent of general relativity. Then, it was
quite common the use of `curvilinear four-dimensional coordinate
system' and `frame of reference' interchangeably, as Bergmann
explains in \cite[pp.158,159]{berg1}: ``\ldots we have always
represented frames of reference by coordinate systems \ldots".
This point is even stressed when he adds: ``The equivalence of all
frames of reference must be represented by the equivalence of all
coordinate systems."

In our discussion, we find necessary to make a distinction between
``reference frames" and ``coordinate systems". By a reference
frame we shall mean an observer set by which measurements are
directly made. For example, a set of radially moving geodesic
observers would comprise a frame of reference. On the other hand,
a coordinate system refers to a set of numbers assigned to each
point in the space-time manifold. That is, we follow a common view
in which ``\ldots coordinates charts are today given a quite
different mathematical status than that of the frames of
reference" \cite[pp.419-434]{EStudies}, whereas they were
previously considered suitable for a given reference frame rather
than for an extended view of the whole manifold.

In Newtonian physics a reference frame is an imagined extension of
a rigid body and a clock. We can then choose different
geometrical coordinate systems or charts (Cartesian, spherical,
etc.) for the same frame. For example, the earth determines a
rigid frame throughout all space, consisting of all points which
remain at rest relative to the earth and to each other. One
can associate an orthogonal Cartesian coordinate system with
such a reference frame in many ways, by choosing three
mutually orthogonal planes and using the coordinates $x, y, z$ as
the measured distances from these planes. As soon as a time
coordinate $t$ is defined one is ready to label any physical event.
It should be stressed that this choice of coordinates presupposes
that the geometry in such a frame is Euclidean.

But what is precisely a reference frame in general relativity? And
how does it differ from a special relativity inertial frame? To
build a physical reference frame in general relativity it is
necessary to replace the rigid body by a fluid
\cite[p.268]{moller} or a cloud of point particles that move
without collisions but otherwise arbitrarily. In more mathematical
terms, one can define \cite[p.627]{fronov} a {\em reference frame}
as a future-pointing, timelike congruence, that is, a
three-parameter family of curves $x^a(\lambda, y^i)$, where
$\lambda$ is a parameter along the curve and $y^i$ is a set of
parameters that `labels' the curves, such that one and only one
curve of the family passes each point. If specific parameters
$\lambda$ and $y^i$ are chosen on the congruence, we define a {\em
coordinate system}. Of course, this choice is not unique. Thus, in
general, a given reference frame can give rise to more than one
associated coordinate system. And a particular coordinate system
may or may not be associated with an obvious reference frame.

Let us define an {\em observer} in a space-time as a material
particle parameterized by proper time \cite[p.36]{o'neill}. An
observer field (or reference frame) on a space-time $M$ is a
future-pointing, timelike unit vector field. Observers enjoy
sending and receiving messages, and keep close track of their
proper time. In special relativity a single geodesic observer can
impose his (or her) proper time on the entire Minkowski
space-time, but in general relativity, ``a single observer is so
local that only cooperation between observers gives sufficient
information"\cite[p.52]{Sachs}, that is, a whole family of
observers is needed for analogous results.

\section{Generalized Observers}

Given the four-velocity field, $u$, of an observer set ${\cal O}$
we parametrize the world lines of ${\cal O}$ with the proper time measured
by a clock comoving with each observer (``wrist-watch time"),
so that we have $g_{ab}u^au^b=u^au_a=-1$; $u^a$ is a geodesic reference
frame iff in addition it is parallel propagated along itself: $\nabla_u u=0$.
The integral curves of $u$ are called observers in $u$ (or $u$-observers, for
short). All observers in a geodesic reference frame are freely falling.

An observer field $u$ on $M$ is {\em stationary} provided that exists a
smooth function $f>0$ on $M$ such that $fu=\xi$ is a Killing vector field,
that is, the Lie derivative of the metric with respect to the vector field
$\xi$ vanishes
\beqa
 L_\xi g_{ab}&\equiv &\xi^c\partial_c g_{ab}+g_{cb}\partial_a\xi^c+
g_{ac}\partial_b\xi^c \nonumber \\
&=&\nabla_a\xi_b+\nabla_b\xi_a=0.
\eeqa
If the one-form corresponding to $\xi$ is also hypersurface orthogonal
$$
\xi_a\equiv \lambda \partial_a\phi,
$$
where $\lambda$ and $\phi$ are two scalar fields, then each $u$-observer is
{\em static} (i.e., $u^\perp$ is integrable). In this case the integral
manifolds $u^\perp$ are three-dimensional, spacelike submanifolds that are
isometric under the flow and constitute a common rest space for the $u$-observers.

Let us consider a test particle given by its 4-velocity vector
field $t^a=dx^a/d\tau$. We can decompose $t^a$ into a timelike
component and a spacelike component by applying a time-projection
tensor, $-(u_au_b)$, and a space-projection tensor, $h_{ab}\equiv
g_{ab}+u_au_b$: \beq t^a_\| =-u^au_bt^b, \quad t^a_\bot =
h^a_bt^b.
 \eeq
 One can easily verify that $t^a_\|$ e $t^a_\bot$
are timelike and spacelike, respectively. Then we rewrite the
space-time distance $ds^2$ between two events $x^a$ and
$x^a+dx^a$ of the test particle's wordline as
\begin{equation}
ds^2=-(u_adx^a)^2+h_{ab}dx^adx^b =-dt^2_\ast + d\ell^2_\ast.
\end{equation}
That is, separation of time and space is always possible
infinitesimally, and an (instantaneous) observer in $x^a$, with four-velocity
$u^a$, measures between the two events $x^a$ and $x^a+dx^a$
of the particle's wordline a proper space and proper time
given respectively by
\begin{equation}
d\ell_\ast =(h_{ab}dx^adx^b)^{1/2},
\label{eq:space}
\end{equation}
and
\begin{equation}
dt_\ast =-u_adx^a.
\label{eq:time}
\end{equation}
The asterisks in Eqs.(\ref{eq:space}) and (\ref{eq:time}) denote
that the quantities so indicated are not, in general, exact
differentials. The minus sign in Eq.(\ref{eq:time}) gives
$dt_\ast$ the same sense as $dx^0$.

There is a natural way for an $u$-observer to define the speed of
any particle with four-velocity $t^a$ as it passes through an
event $p\in M$. As the observer has instantaneous information at
$p$ that allows him (or her) to break up the tangent space
$T_p(M)$ at $p$ into time $t$ (parallel to $u$) and space
$u^\bot$ , he (or she) will measure
\begin{equation}
v^2=(\frac{d\ell_\ast}{dt_\ast})^2=\frac{(g_{ab}+u_au_b)t^at^b}{(u_at^a)^2},
\label{veloc5}
\end{equation}
for the square of the speed of the particle at $p$, which can be
written as,
\begin{equation}
v^2=1-\frac{1}{(u_at^a)^2}.
\label{veloc2}
\end{equation}

Whatever is the particle's four-velocity, $t^a$, one can always
write it as \beq t^a=t^a_\|+t^a_\bot=\lambda u^a+\ell^a, \quad
\mbox{where} \quad \ell_a u^a=0. \eeq Since $t^a$ should be
timelike, $t^at_a=-\lambda^2+|\ell|^2<0$ (notice that
$|\ell|^2=\ell^a \ell_a=h_{ab}t^at^b=(t^a)_\bot (t_a)_\bot)$, and
since both $t^a$ and $u^a$ are future-pointing, $\lambda=-u_at^a
>0$, and $|\ell|<\lambda=(|\ell|^2+1)^{1/2}$. From this last
equation and from (\ref{veloc2}) one immediately concludes that
under these conditions $v^2<1$. The number $\lambda$ represents
the instantaneous rate at which the observer's time is increasing
relative to the particle's time, and $|\ell|$ is the rate at which
arc length $d\ell_\ast$ in $u^a_\bot$ is increasing relative to
the particle's time, that is, \beq \lambda=\frac{dt_\ast}{d\tau},
\quad |\ell| =\frac{d\ell_\ast}{d\tau}. \eeq

Thus the $u$-observer measures the speed of the $t$-particle at event $p$ as
\beq
v=\frac{d\ell_\ast}{dt_\ast}=\frac{d\ell_\ast/d\tau}{dt_\ast/d\tau}=
\frac{|\ell|}{\lambda}<1.
\label{veloc3}
\eeq
Notice that, from Eq. (\ref{veloc3}), $v=1$ iff the $t$-particle is
lightlike ($t^at_a=0$); otherwise, for timelike particles, $v<1$.

\section{The Schwarzschild Field Case}

Having dealt with this problem in a very general way and proved
that the velocity $v$ of any massive particle with respect to any
physical observer is always smaller than the velocity of light:
$v<1$, let us apply these ideas to the Schwarzschild
gravitational field and find a general prescription for
evaluating $v$ when both the particle and the observer are
geodesic.

\subsection{Geodesic test particle}

Let us suppose that our test particle follows a radially ingoing
geodesic in a Schwarzschild field. Its geodesic equation of motion
is the Euler equation for the Lagrangian $2L=g_{ab}\dot x^a\dot
x^b$, which is given by \beq 2L=-\alpha \dot t^2 +\alpha^{-1} \dot
r^2, \label{lagrange} \eeq where
$\alpha=-g_{00}=g_{11}^{-1}=1-2m/r$, for the Schwarzschild metric
Eq.(\ref{metschw}) with $\theta=const.$ and $\varphi=const.$, and
the dot, as usual, denotes differentiation with respect to proper
time. Along the orbit \beq 2L=-1, \label{lagrange1} \eeq for the
particle's proper time is given by \beq d\tau^2=\alpha  dt^2
-\alpha^{-1} dr^2. \label{radial_m} \eeq From this we could also
write \beq d\tau^2=\alpha dt^2(1-v^2), \label{veloc0} \eeq where
\beq v^2=\frac{1}{\alpha^2}\left(\frac{dr}{dt}\right)^2,
\label{veloc1} \eeq is, accordingly to Eq.(\ref{veloc5}), the
velocity of the particle with respect to a static observer
($r=constant$); i.e. while the particle travels a proper distance
${\alpha}^{-1/2}dr$ the observer measure a proper time given by
${\alpha}^{1/2}dt$.

Eq. (\ref{lagrange}) shows that $t$ is a cyclic coordinate, and
\beq
-\frac{\partial L}{\partial \dot t}=(1-2m/r)\dot t=const.=: E,
\label{energy}
\eeq
is the constant of motion along the geodesic associated with the Killing vector
field $\partial /\partial t$; that is, if the particle's 4-velocity $t^a$ is
geodesic, $\nabla_t t=0$, then: $\nabla_t [g(t, \partial /\partial t)]=0$, which
equally implies Eq. (\ref{energy}).

Inserting Eqs.(\ref{veloc1}) and (\ref{energy}) into
Eq.(\ref{lagrange1}) gives \beq
\left(\frac{dr}{d\tau}\right)^2=E^2-\alpha, \label{radial2} \eeq
and from this we obtain \beq
E^2=\frac{\alpha}{1-v^2}=\frac{1-2m/R}{1-v_0^2}, \label{energy2}
\eeq where $(R, v_0)$ are initial conditions; $R$ is the radial
coordinate at which the fall begins, and $v_0$ is the initial
velocity.

Now, from Eqs.(\ref{energy}) and (\ref{radial2}) we obtain the
components of the 4-velocity $t^a$ of a radially ingoing geodesic
particle \beq t^a=\left(\frac{E}{\alpha}, -\sqrt{E^2-\alpha}, 0,
0\right), \label{4veloc} \eeq written in terms of its constant of
motion $E$.

\subsection{Static limit}

In Landau and Lifchitz \cite[p.250]{landau} the velocity is
measured in terms of proper time, as determined by clocks
synchronized along the trajectory of the particle, as they say.
Their prescription leads to the following expression
 \beq
v^2=\left(g_{00}+g_{01}\frac{dx^1}{dx^0}\right)^{-2}
\left(g_{01}^2-g_{00}g_{11}\right)(\frac{dx^1}{dx^0})^2,
\label{velocll} \eeq for the square of the velocity of a radially
moving particle.

We have seen earlier, that there is a natural way for the
observer $u^a$ to measure the speed of any particle with
four-velocity $t^a$ as it passes through an event $p\in M$, which
is coordinate free, and given by Eq.(\ref{veloc2}). For a static
observer the 4-velocity has the following components
$$u^au_a=-1\Rightarrow u_a=\left(-g_{00}\right)^{-1/2}g_{a 0},$$
and for the test particle, its tangent vector to radially inward,
timelike geodesics may be written as
$$
t^a=\left(\frac{dx^0}{d\tau}, \frac{dx^1}{d\tau}, 0, 0\right).
$$
Inserting these last two 4-vector components in Eq.(\ref{veloc2})
leads to Eq.(\ref{velocll}), which must be understood as a
specialization of Eq.(\ref{veloc2}) for static observers.

 When applied to the (geodesic) radial motion of a
free falling particle, Eq.(\ref{velocll}) leads to
Eq.(\ref{veloc1}) which can be rewritten as
 \beq
v=\left[1-\frac{1-2m/r}{E^2} \right]^{1/2},
 \label{veloc2a} \eeq
which is equivalent to Eq.(\ref{velfn}). In the case when $E=1$,
corresponding to $R=\infty$ or $v_0=0$, it reduces to \beq
v=\left({{2m}\over {r}}\right)^{1/2}, \label{veloc3a} \eeq which
coincides with the Newtonian expression

For either expression, $v$ approaches the speed of light at the
event horizon ($r=2m$) and they seem to to predict
faster-than-light speeds inside the black hole\cite{sabbata}. It
is easily seen that \beq \lim_{r\rightarrow 2m} v=1 \quad
\mbox{and} \quad \lim_{r\rightarrow 0} v=\infty, \eeq for both
Eqs. (\ref{veloc2a}) and (\ref{veloc3a}).

Taken at face value the previous statements would imply that the
particle's trajectory should become lightlike in the limit
$r\rightarrow 2m$. However, as the trajectory can be continued
through the event horizon, it seems clear that it must remain
timelike there, otherwise we had to conclude that the particle's
velocity would overcome the light speed as its worldline becomes
spacelike.

Since this is an unacceptable result and we know that the
Schwarzschild coordinate system is not suitable for describing the
manifold at $r=2m$ it is rather tempting to blame the coordinate
system for this malfunction. But we should ask first, could it be
possible to find a coordinate system that does not have this
defect? The answer is obviously no, since the result is
independent of the choice of coordinates, as we have proved in the
third section of this paper, and as will be even clearer at end of
this section. Indeed, even if we use a coordinate system that has
no difficulties at $r=2m$, like the advanced Eddington-Finkelstein
coordinates, we would still end up with the same result
$v^2\rightarrow 1$ as $r \rightarrow 2m$.

We can easily see this by introducing the Eddington-Finkelstein
metric \cite[p.828]{mtw},
\begin{equation}
\label{ef}
 ds^2=-\left( {1-{{2m} \over r}} \right)dw^2+2dwdr+r^2\left(
{d\theta ^2+\sin ^2\theta d\varphi ^2} \right),
\end{equation}
where
\begin{equation}
 w(t,r)=t+r+2m\ln \left| {{{r-2m} \over {2m}}} \right|,
\end{equation}
in Eq.(\ref{velocll}), valid for static observers. Then, at
$r=2m$, where $g_{00}=0$, we obtain $v^2=1$.

Thus the real issue here is the choice of frame not the choice of
coordinates. For instance, the process of synchronizing clocks,
used by Landau and Lifchitz\cite[p.250]{landau}, involves the
determination of simultaneous events at different spatial
locations, which is a frame-dependent prescription.

Notice also that an observer cannot stay at rest in a Schwarzschild field
at $r=2m$, where $g_{ab}u^au^b=0$, for he (or she) cannot have there a
timelike four-velocity field tangent to its worldline.
This means that only a photon can stay at rest at $r=2m$, and with respect to
this ``photon-frame" all particles have $v^2=1$, as it should be expected.

Another argument that could be given, although it is closely
related to the later discussion, is provided by the study of the
acceleration of a static observer in a Schwarzschild field
\cite{dough}. Whereas a static Newtonian observer is considered to
be at rest in its own proper ``inertial frame", in general
relativity an observer at rest is not geodesic and is accelerated.
To make it clear(er) let's evaluate the acceleration of a static
observer in a spherically symmetric and static gravitational
field. Starting with its four-acceleration field components in
Schwarzschild coordinates,
$$
a^c=u^c_{\; ;\;b}u^b=u^c_{\;,\;b}+u^a\Gamma_{ab}^c=g^{ca}g_{00,\;a}g_{00}^{-1}/2,
$$
one finds that all nonradial components vanish \beq
a^0=a^\theta=a^\varphi=0, \; \mbox{and} \; a^r=-\frac
12g_{00,1}=\frac{m}{r^2}. \eeq However, $a^r$, which is the
radially inward acceleration as calculated using Newtonian
gravity, is a coordinate-dependent quantity and it is not a scalar
field. The invariant acceleration magnitude that we require is
$$
a=(a^ca_c)^{1/2}=-(g_{11})^{1/2}g_{00,1}/2.
$$
For the Schwarzschild field this gives
\beq
a=\frac{m}{r^2}\left(1-\frac{2m}{r}\right)^{-1/2}.
\eeq
The factor $(g_{11})^{1/2}$, by which the GR and Newtonian accelerations differ,
can be neglected in most cases ($r\gg 2m$), such as apply e.g. on the
surface
of planets or even on the surfaces of normal stars. For instance, for the
Sun $(2m/R)_\bigodot=4.233\times10^{-6}$. But on the surface of a neutron
star $(g_{11})^{1/2}$ may exceed unity by a very large factor, and for a
black hole
$$
a \rightarrow \infty, \quad \mbox{as} \quad r\rightarrow 2m.
$$
``It follows that a `particle' at rest in the space at $r=2m$
would have to be a photon" \cite[p.149]{Rind1}. This makes it very
clear that we should define a {\em static limit} of a black hole,
that is, the boundary of the region of space-time in which the
observer can remain at rest relative to any observer in the
asymptotically flat space-time. In plain words, as any observer
must follow a timelike worldline, the static limit is given by
$$
g_{00}(r)=0 \quad \mbox{(static limit)}.
$$
This emphasizes the point that one cannot use expressions like
(\ref{veloc1}) or (\ref{velocll}) at the surface $r=2m$. In other
words, there is no observer at rest on that surface. As Taylor and
Wheeler put it in their recent textbook \cite[p.3-15]{T&W}:
``Shell--and shell observers--cannot exist inside the horizon or
even {\em at} the horizon, where the spherical shells experiences
infinite stresses."

\subsection{Radial observers}

Considering that the particle and the observer are both in free
fall (inward, timelike geodesics), we can use Eq.(\ref{4veloc})
and write respectively \beq t^a=\left(\frac{ E_1}{ \alpha},
-\sqrt{E_1^2-\alpha}, 0, 0\right), \quad u^a=\left(\frac{E_2}{
\alpha}, -\sqrt{E_2^2-\alpha}, 0, 0\right). \label{4veloc2} \eeq
Then inserting these into Eq.(\ref{veloc2}) the following
expression is obtained, \beq v^2=1-\frac{\alpha^2}{E_1^2
E_2^2\left[1-\sqrt{1-\frac{\alpha}{E_1^2}}
\sqrt{1-\frac{\alpha}{E_2^2}}\right]^2}, \eeq and since
$\alpha=1-2m/r,$ it follows that
$$ \lim_{r\rightarrow 2m} v^2=1-0/0.$$
However, $(1-\alpha/ E^2)^{1/2}$ may be expanded if $r\approx 2m$
since $\alpha/ E^2\ll 1$,
$$
\left(1-\frac{\alpha}{ E^2}\right)^{1/2}=1-\frac 12\frac{\alpha}{
E^2}-\frac 18\frac{\alpha^2}{ E^4} +O\left(\frac{\alpha}{
E^2}\right)^3.
$$
This leads to \beq
 \label{formbef}
 v^2=1-\frac{\alpha^2}{E_1^2E_2^2}\left[\left(\frac{\alpha (E_1^2+E_2^2)}{2E_1^2E_2^2}
 +O\left(\frac{\alpha}{
E^2}\right)^2\right)^2\right]^{-1}
 \eeq
 and we now obtain an exact expression for the velocity at
$r=2m$, \beq
 v^2(r=2m)=1-\frac{4 E_2^2 E_1^2} {(E_2^2+E_1^2)^2}
\label{formula} \eeq which shows (see figure1) that the value of
the velocity at $r=2m$ is smaller than 1 unless either $E_1$ or
$E_2$ are zero or infinity. In particular, when $E_1=E_2$, we see
that $v^2(r=2m)=0$. This means that particle and observer have the
same initial conditions at some space-time point $p$ with $r>2m$,
and from that event onwards they are both on the same local
inertial frame.

\begin{figure}
\includegraphics[width=8cm]{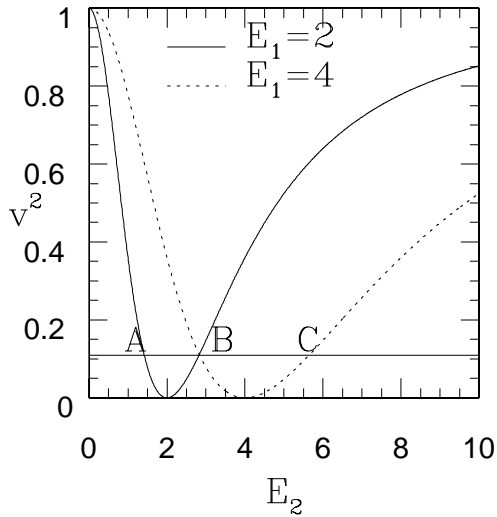}
\caption{Test particle's square speed $v^2$ at $r=2m$ as function
of the observer's constant of motion, for two cases (particle with
$E_1=2$ and $E_1=4$).}
\end{figure}

Notice that for each particle there are 2 observers who measure
the same value of $v^2$. For example observers A and B for one
particle and B and C for the other. Notice also that the constant
of motion of the particle is always in between the two values of
$E_2$ of those observers. This means, picturing the first particle
and the observers A and B all free falling, that when they all
meet at $r=2m$, the particle reaches the observer A with the same
velocity that the observer B reaches the particle with(and the
same for B,C and the other particle). In other words, the value of
$v$ is positive in the first branch of the plots (before the
minimum) and negative in the second branch.

Considering now the 2 particles, we notice there is an observer
(B), with $E$ in between the values of the particles' constants of
motion, who measures the same value of $v^2$ for both particles.
Once again this means that at $r=2m$, B catches one of the
particles (the first one) with the same speed with which it is
caught by the other.

In fact, owing to symmetry between test particle and observer,
both situations are equivalent. And we could say that figure 1
refers to the observer's square speed $v^2$ at $r=2m$ as function
of the particle's constant of motion.

Now, we highlight the fact that only in the limits $E\rightarrow
0$ and $E\rightarrow \infty$ (from what we have just seen, it is
indifferent if $E$ refers to the particle or to the observer), we
obtain $v=1$. In these cases we conclude the hypothetical
`observer' (or test particle) is in a ``photon-frame". In fact
referring to Eq.(\ref{energy2}) we see these two limits correspond
to either $v_0=1$ or $R=2m$. With respect to this (unphysical)
frame all particles travel at the speed of light $v=1$.

Expressions similar to Eq.(\ref{formula}) can be found
\cite{janis1,tereno} for the velocity of a free falling particle
in the Schwarzschild field, derived for diverse non-static
observers.

As an example, let us consider a Kruskal observer, an observer
which follows an orbit defined by
$$u^a=\left( {{dt'}\over{d\tau}},{{dx'}\over{d\tau}}\right),$$
with $dx'=0$, where $(t',x')$ are the Kruskal coordinates.

For $r>2m$, these coordinates $(x',t')$ relate to the
Schwarzschild ones by,
\begin{equation}
\label{xtext} \left\{
\begin{array}{lll}
\mbox{$x'^2-t'^2=\left( {{{r-2m} \over {2m}}} \right)e^{r/2m}$} \\ \\
\mbox{$t'=\tanh \left( {{t \over {4m}}} \right)x'$}
\end{array}
\right.
\end{equation}
and the metric takes the form \cite[p.832]{mtw},
\begin{equation}
\label{kruskal} ds^2={{32m^3}\over
{r}}e^{-r/2m}(-dt'^2+dx'^2)+r^2\left( {d\theta ^2+\sin ^2\theta
d\varphi ^2} \right).
\end{equation}
From here we see that an observer which maintains the space-like
coordinate $x'$ constant, verifies,
\begin{equation}
\label{movkru} {{32m^3} \over {re^{r/2m}}}\left( {{{dt'} \over
{d\tau }}} \right)^2=1.
\end{equation}
Differentiating Eq.(\ref{xtext}) we get,
\begin{eqnarray}
\label{ddtau} {{{dr} \over {d\tau }}}={{8m^2} \over {e^{r/2m}r}}
\left( x'{{{dx'} \over {d\tau }}} -t' {{{dt'} \over {d\tau
}}}\right), & & {{dt} \over {d\tau }}=\left( {x'{{dt'} \over
{d\tau }}-t'{{dx'} \over {d\tau }}} \right){{8m^2} \over
{e^{r/2m}(r-2m)}}.
\end{eqnarray}
Using $dx'=0$ and Eq.(\ref{movkru}) we can write the following
equation :
\begin{equation}
\left( {1-{{2m} \over r}} \right)\left( {{{dt} \over {d\tau }}}
\right)^2 -\left( {1-{{2m} \over r}} \right)^{-1}\left( {{{dr}
\over {d\tau }}} \right)^2={{2m(x^2-t^2)} \over
{e^{r/2m}(r-2m)}}=1,
\end{equation}
which shows this observer follows a radial trajectory.

Consider now a material particle along a radial ingoing geodesic.
From Eq.(\ref{veloc5}), its velocity, measured by a Kruskal
observer is
\begin{equation}
v={{dx'}\over{dt'}},
\end{equation}
since Eq.(\ref{kruskal}) is diagonal with $g_{x'x'}=g_{t't'}$,
which is analogous to Eq.(\ref{veloc1}). Dividing one of the
equations Eq.(\ref{ddtau}) by the other and solving for $v$ we
obtain,
\begin{equation}
\label{v} v={{1+\tanh(t/4m){{dt}\over{dr}}(1-2m/r)}
\over{\tanh(t/4m)+{{dt}\over{dr}}(1-2m/r)}},
\end{equation}
where $dt$ and $dr$ refer to the movement $t(r)$ of the particle.

We can now introduce the geodesic $dt/dr$ followed by the
particle, from Eq.(\ref{radial2}) which we can also explicitly
integrate to obtain an expression for $t(r)$ to substitute in
Eq.(\ref{v}). The details can be found in \cite{tereno}, where the
behavior of the $v$ against $E$ plot was found to be identical to
the one presented in this paper.

\section{Conclusions and Discussion}

We have seen that the speed of any material particle following a
radially inward geodesic is strictly less than 1 with respect to
any physical (timelike) observer. We recalled that there is a
limit for the use of static observers in a Schwarzschild field
given by: $g_{00}(r)=0$. Thus, we stress the point that one can
only use static observers in the space-time region characterized
by $r>2m$. We found a formula for the physical velocity of a test
particle in a radially inward, timelike geodesic, measured by an
observer in free fall (which crosses the event horizon
simultaneously with the particle) valid at $r=2m$.

We conclude that all free falling observers crossing the black
hole surface measure the speed of light (`standing still' photons
at $r=2m$) to be $v=1$, and they measure the speed of any
material particle to be strictly less than 1.

\section{Acknowledgments}
During the last part of this work, one of us (PC) visited the
Physics Department of the University of Boston and the Center for
Einstein Studies, and he doubly thanks that university, and in
particular John Stachel, for hospitality. He also thanks J.
Stachel for a critical reading of an earlier version of the
manuscript and for bringing some references to his attention. We
also thank Rosa Doran for many stimulating discussions on this and
related issues.

\end{document}